\documentclass{article}
\usepackage{spconf,amsmath,graphicx}
\usepackage{amsfonts}
\usepackage[]{algorithm2e}
\usepackage{algorithmic}
\usepackage{url}
\usepackage{cite}
\usepackage{draftwatermark}
\SetWatermarkText{DRAFT}
\SetWatermarkScale{0.5}

\title{A Pairwise Approach to Simultaneous Onset/Offset Detection\\ for Singing Voice using Correntropy}
\name{Sungkyun Chang and Kyogu Lee\thanks{This research was supported by Samsung Electronics and the MSIP (Ministry of Science, ICT \& Future Planning), Korea, under the ITRC (Information Technology Research Center) support program supervised by the National IT Industry Promotion Agency. NIPA-2013-H0301-13-4005.}}
\address{Music and Audio Research Group\\ Seoul National University, 151-742 Seoul, Korea\\
e-mail:\{rayno1,kglee\}@snu.ac.kr}
\begin{document}

\maketitle
\begin{abstract}
In this paper, we propose a novel method to search for precise locations of paired note onset and offset in a singing voice signal. In comparison with the existing onset detection algorithms, our approach differs in two key respects. First, we employ {\it Correntropy}, a generalized correlation function inspired from {\it Reyni's entropy}, as a detection function to capture the instantaneous flux while preserving insensitiveness to outliers. Next, a novel peak picking algorithm is specially designed for this detection function. By calculating the fitness of a pre-defined inverse hyperbolic kernel to a detection function, it is possible to find an onset and its corresponding offset simultaneously. Experimental results show that the proposed method achieves performance significantly better than or comparable to other {\it state-of-the-art} techniques for onset detection in singing voice. \end{abstract}
%

\section{Introduction}
\label{sec:intro}

Onset detection is a problem of finding the precise location of discrete musical events, and thus is an important pre-processing step in many music applications, including pitch estimation, beat tracking, and automatic music transcription, to name a few. In music signal processing, note onset detection still remains a challenging problem, particularly for singing voice,  because of several reasons such as a large variance in articulation, singer-dependent timbral characteristics, and slowly-varying onset envelopes. The report from {\it Music Information Retrieval Evaluation eXchange} 2012 (MIREX 2012) for solo singing voice reflects these difficulties, where the best-performing algorithm yields the F-measure of merely 55.9\% \cite{mirex2012perclass}. It is noteworthy that singing class gives much lower F-measure value than other classes like polyphonic pitched, solo brass, and wind instruments classes. 

Numerous methods have been proposed so far to solve the onset detection problem \cite{Bello}. Majority of existing methods are generalized by the two-stage procedures: detection function and peak picking \cite{Bello,Dixon,Duxbury,Heo,Toh,Eyben,Bock,Wang}. Generating detection function refers to transforming audio signals into feature vectors more relevant to indicating onsets. Due to the fact that singing voice contains many soft onsets \cite{Bello}, a recent study focuses on finding the harmonic regularity instead of using the conventional, energy-based techniques \cite{Heo}. If a properly designed detection function is provided, onset candidates will give a rise to certain local peaks. The peak-picking procedure then finds precise onset locations from the detection function.

\begin{figure}
	\centerline{{
	\includegraphics[width=1\columnwidth]{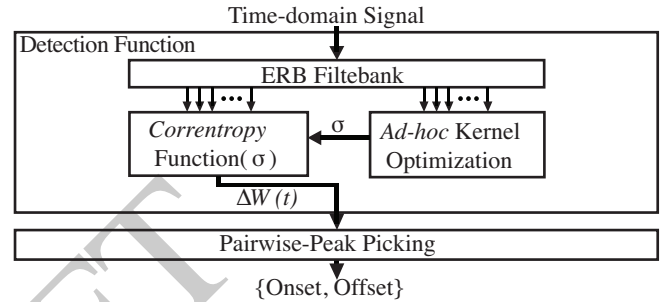}}}
	\caption{Overview of the system}
 	\label{fig:system}
\end{figure}


This paper explores a new detection function and a peak-picking method for onset/offset detection in singing voice. The use of correntropy for detection function brings us two major advantages: first, it provides a compact front-end like a conventional correlation function; second, the property of correntropy preserves robustness to outliers -- such as noise or subdominant changes in frequency, amplitude or phase -- by projecting an input signal into a high dimensional Reproducing Kernel Hilbert Space (RKHS).

Figure \ref{fig:system} illustrates an overview of the proposed system. We describe a novel detection function based on correntropy in Section 2.  An \emph{ad-hoc} kernel optimizer continuously updates a localized parameter required for correntropy estimation.  The detection function yields a regular shape with distinct local peaks, or ($-$)/($+$) peaks, which correspond to onset/offset, respectively. This regular shape motivates us to design a pairwise peak-picking method in Section 3. The experimental results are presented in Section 4, followed by concluding remarks in Section 5. A relation to prior work is addressed in Section 6.

 


\section{Detection Function based on Correntropy}\label{sec:typeset_text}
Our detection function first passes the time-domain input signal sampled at 11,025 Hz into an auditory filterbank\cite{moore}. We map center frequencies of 64-channel gammatone filterbank according to the {\it Equivalent Rectangular Bandwidth} (ERB) scale between 80 Hz and 4,000 Hz. Hereafter,  $x_c(t)$ refers to the amplitude of the filtered parallel output, where $c$ and $t$ denote the channel and discrete time indices, respectively.


\subsection{Correntropy-based Formulation}

\begin{figure}[t]
	\centerline{{
	\includegraphics[width=1\linewidth]{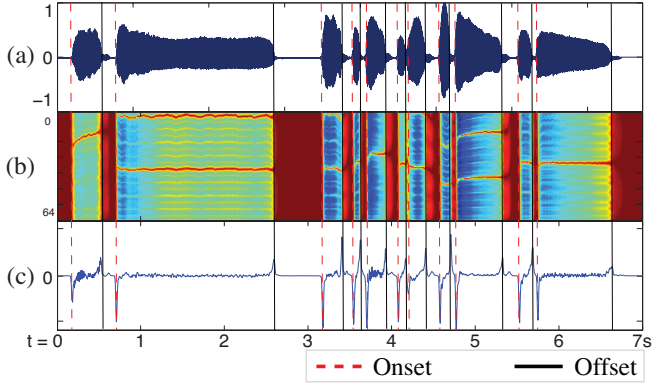}}}
	\caption{Singing voice example. From top to bottom: (a),(b) and (c) are input signal,  $W_{t}(\tau)$ and detection function $\Delta W(t)$, respectively. Red dashed-lines are onset positions that correspond to (-) peaks in (c). Black solid-lines are (+) peaks near the offset. }
 	\label{fig:exam1}
\end{figure}
Correntropy function incorporates both distribution and temporal structures of a time series in random processes \cite{Liu}.  There are various applications of correntropy to non-linear and non-gaussian signal processing where correlation function is not sufficient. Let \begin{math} \{x_t : t \in T\} \end{math} be a random process with \begin{math}T\end{math} denoting an index set. Correntropy function  \begin{math}V\end{math} is defined as 
\begin{equation}
V(x_{t_1},x_{t_2}) = \mathbb{E}[\kappa(x_{t_1},x_{t_2}|\sigma)],
 \end{equation}
where \begin{math}\kappa_{\sigma}(\cdot)\end{math} is a Parzen kernel  and \begin{math}\mathbb{E} \end{math} is an expectation operator. $V$ is advanced to measure the distance between the two discrete vectors.

 The properties reveal that correntropy has very similar characteristics to a conventional correlation function. Liu \emph{et al.} showed that the sufficient condition to satisfy \begin{math}V(t,t-\tau) = V(\tau)\end{math} is that the input random process must be time shift-invariant on the even moments. More specifically, this is a stronger condition than a wide-sense stationarity involving only second-order moments. We estimate correntropy $V$  given $t$, $c$ and lag $\tau$  by computing the sample mean of a size $N$ window as follows:
\begin{equation}
V_{t,c}(\tau)=\frac{1}{N}\sum_{n=1}^{N}\mathcal{N}_{\sigma}(x_c(t+n),x_c(t+n+\tau)),
\label{vtc}
\end{equation}
where \begin{math} \mathcal{N}_\sigma(p,q) = \frac{1}{\sqrt{2\pi}\sigma} \text{exp}\{-\frac{(p-q)^2}{2\sigma^2}\} \end{math}, and $\sigma$ is a Gaussian bandwidth parameter.
In practice, both the window size $N$ and maximum $\tau$ are set to $\text{sampling rate}/80$, when the lowest band-limit is 80  Hz. Then the correntropy coefficient for each channel is ``pooled'' into a non-negative summary matrix $W$ as follows \cite{Xu}:
\begin{equation}
W_t(\tau)=\sum_{c}V_{t,c}(\tau).
\label{wt_tau}
\end{equation}
The detection function $\Delta{W(t)}$ is then calculated by the rectified difference with a hop-size $h$ as follows:
\begin{equation}
\Delta W(t) = \sum_{\tau}W_{t+h}(\tau)-\sum_{\tau}W_{t}(\tau).
\label{dwt}
\end{equation}

A monophonic singing example in Figure \ref{fig:exam1} illustrates onset/offset regions and their relation to the detection function. We estimate onset/offset locations by finding the first time $t$ at which $\Delta{W(t)}<0$, and the first time after $t$ where $\Delta{W(t)}>0$, respectively. 

\subsection{{\it Ad hoc} Kernel Optimization}

\begin{figure}[!t]
	\centerline{{
	\includegraphics[width=1\linewidth, ]{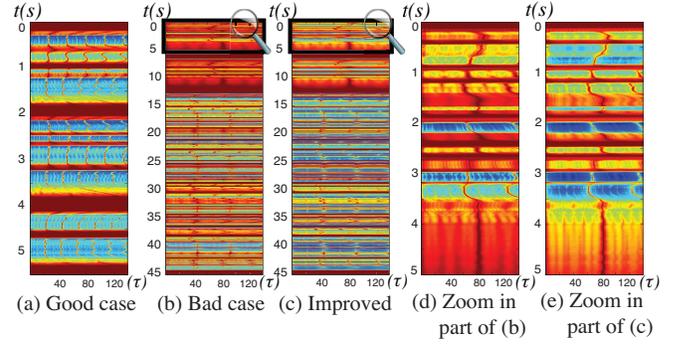}}}
	\caption{ Examples of $W_{t} (\tau)$ in Section 2.2: (a) $\sigma=0.017$ preserves good contrast between stationary and non-stationary regions. (b) bad case using global optimum for 45 s. (c) improved result with ``loosely'' localized $\sigma$. (d), (e) zoomed-in parts from (b) and (c). We observe more contrast in (e).}
 	\label{fig:exam2}
\end{figure}
\begin{figure*}[!t]
	\centerline{{
	\includegraphics[width=1\linewidth]{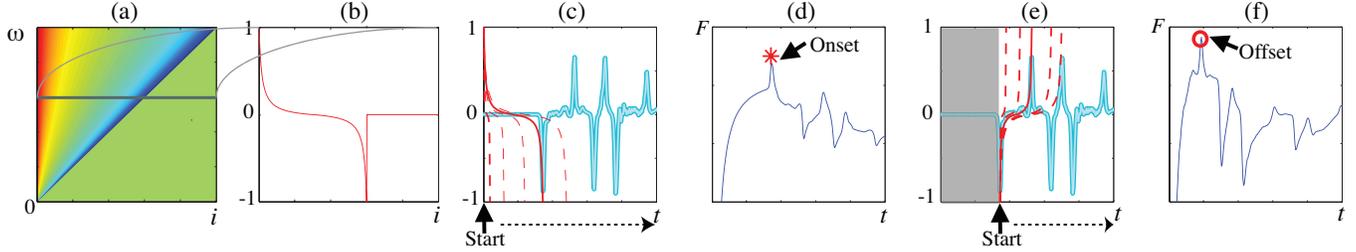}}}
	\caption{Illustration of peak-picking algorithm. (a) Pre-defined kernel matrix $\Lambda$. (b) Cross section of  $\Lambda$ with $\omega$. In (c), start search at $t=0$. Calculate fitness by stretching $\Lambda$. In (d), onset is found at maximum fitness location. In (e), start search from previous onset location using $-\Lambda$. In (f), corresponding offset is found similarly as in (d). We repeat (c) to (f) until the end.}
 	\label{fig:peakpicking1}
\end{figure*}


A free parameter $\sigma$ in Equation \eqref{vtc} acts as a sensitivity controller for detection function: the larger \begin{math}{\sigma}\end{math} becomes, the faster the higher-order moments decay. To keep both nonlinearity and discrimination ability, one way to select the optimal $\sigma$ is given by Silverman's {\it Rule of Thumb}\cite{silverman}. Assuming that $\kappa$ is a  Gaussian kernel, the optimal $\sigma$ can be simplified to

\begin{equation}
\sigma = b\cdot\hat{\psi}N^{-1/5},
\label{silvermaneq}
\end{equation}
where b, $\hat{\psi}$ and $N$ denote the constant scale factor, sample standard deviation, and the number of samples, respectively.

The sensitivity of the detection function can be observed from the contrast in colors in $W_{t}(\tau)$. Figure \ref{fig:exam2}(a) is a good example where the boundaries would yield a relevant detection function using the global optimum $\sigma$. However, Figure \ref{fig:exam2}(b) reveals that the global optimum parameter doest not always guarantee good contrast on the entire song. On the other hand, a strongly localized parameter would not guarantee temporal contrast. In this situation, a ``loosely'' localized optimal parameter is required for robust detection of real-world singing onsets. Figure \ref{fig:exam2}(c) displays $W_{t}(\tau)$ with improved contrast by an {\it ad-hoc} optimizer. It updates $\sigma$ with an interval $h$. In practice, we achieve good performances by computing the optimal $\sigma$ using Equation \eqref{silvermaneq} with an observation window size of 7 s and $h = 5$ ms. This improves the precision by over 10\% in comparison with a global optimization method.

\section{Pairwise Simultaneous Peak Picking}
A pairwise peak-picking approach is motivated by the regular shape of the detection function observed in Figure \ref{fig:exam1}(c): a basic idea is to capture a pair of falling and rising peaks by calculating a fitness to a pre-defined kernel. Figure \ref{fig:peakpicking1} illustrates this concept.
 
We first generate a set of pre-defined inverse hyperbolic kernels, whose shape is similar to expansion or shrinkage of detection function. This kernel $\Lambda$ is defined as
\begin{equation}
\Lambda(z) = \frac{z}{1+\alpha-|z|},
\label{prediefinedk}
\end{equation}
such that $-1+10^{-5}\leq{ z} \leq  1-10^{-5}$, where 
$\alpha$ is a sharpness factor. The range of $z$ is set to avoid division by 0, and $\alpha\approx0.15$ is empirically found. 
Given an observation window length $\omega$ 
and sample index $i=\{1,2,...,\omega\}$,
we chop $z$ into $\omega$ samples by a linear scale. 
In our default settings,  $\omega_{min}$ is set to 4 samples (=20 ms) and $\omega_{max}$ is set to 500 samples or larger ($\geq 2.5$ s) for 5 ms-correntropy hopsize $h$. Hence, any event less than 20 ms will be ignored. Figure \ref{fig:peakpicking1}(a) displays the generated kernel matrix for a set of observation length $\omega$.

To find a pair of offet/onset, we calculate the fitness between the detection function and the pre-defined kernel as we expand the kernel size. The fitness is calculated by 
\begin{equation}
Fit(\Lambda_{\omega}',W_{\omega}') = (\Lambda'_{\omega} - W'_{\omega})^2 \cdot {\omega}^{-k},
\label{eq:fitness}
\end{equation}
where $\Lambda'_\omega$ is the pre-defined kernel sampled at $\omega$, and $W'_\omega$ is the $\omega$-long rectangular windowed detection function. $k$ is a weighting factor for close peaks and we set $k=1$. For reference, Equation \eqref{eq:fitness} is derived from {\it lack-of-fit} sum of squares which has been widely used in classical F-test statistics\cite{ftest}. To find a pair of onset/offset, we start from the last onset position and perform the same calculation but use $-\Lambda$.

The above procedures are summarized in Algorithm 1. This pairwise approach makes sense in monophonic sources: if an onset is found, then its corresponding offset must exist before the next onset.
\begin{algorithm}[]
\SetKwInOut{input}{Input}\SetKwInOut{output}{Output} 

\DontPrintSemicolon

\input \; 
$\omega$ = window size,
$\Lambda$ = inverse hyperbolic kernel. \;
$W$ = detection function.\;
\output \;
 $\star, \circ$ are onset and offset marking on time index, $t$.
\;
 \While{$t\rightarrow T$}{
\%{\it find onset,} $\star$\; 
\For{$t':(t+\omega_{\text{min}})$ $\rightarrow$ $(t+\omega_{\text{max}})$}{
$F(t') = Fit(\Lambda',W')$ 
}
$ \star = t + \arg\max_{t'}{(F(t'))}$ \;{ $t=\star$   }\;
\% {\it find offset,} $\circ$\;
\For{$t':(t+\omega_{\text{min}})$ $\rightarrow$ $(t+\omega_{\text{max}})$}{
$F(t') = Fit(-\Lambda',W')$ 
}
 $ \circ = t + \arg\max_{t'}{(F(t'))}$\; 
 $t=\circ$\;
 }\;
\SetAlgoLined
\label{algo:pseudo1}
 \caption{Pseudo code for pairwise peak picking described in Section 3.}
\end{algorithm}


\section{Experiment}
\subsection{Dataset}
The dataset is obtained from the authors of referenced paper \cite{Heo}. It allows us to directly compare the performance of proposed algorithm to theirs. The total length of audio clips is about 13 minutes, which is much longer than singing data included in the MIREX audio onset detection task\cite{mirex}. The dataset consists of 13 male and 2 female singers' recordings of popular songs. Onset labels are cross-validated by three persons who have professional careers in music. In total, the dataset contains 1,567 onsets with annotations. Audio files are produced in mono with the sampling rate of 44,100 Hz. 

\subsection{Results and Discussion}
\begin{table}[h]
\noindent
	\begin{center}
		\begin{tabular*}{\linewidth}{c c c c c} 
			\hline\hline 
			{\it Class} & {\it \# of Onset} & Precision & Recall & F-measure\\ [1.5ex] 
			\hline 
			male & 1,533 & 80.9 & 80.1 & 80.3\\
			female & 34 &93.8 & 88.4 &91.4\\ [1ex]
			Total & 1,567 &  81.1 & 80.2 & 80.6\\
			\hline 
		\end{tabular*}
	\end{center}
\caption{Performance of the proposed algorithm }
\label{table:result}
\end{table}
\begin{figure}[h]
	\centerline{{
	\includegraphics[width=1\linewidth]{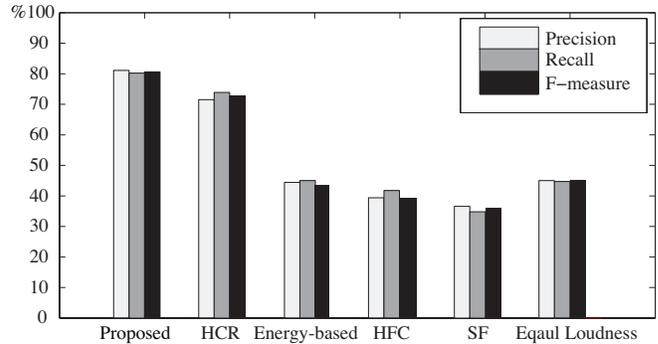}}}
	\caption{Comparison of performance with existing onset detection algorithms. From left to right, the proposed and other algorithms [4, 14, 1, 3, 13].}
 	\label{fig:resultbar}
\end{figure}
We followed the evaluation procedure for onset detection described in MIREX \cite{mirex}. The tolerance value is set to $+/-$ 50 ms, which allows us to consider any detected onset within this range from the ground-truth onset as a true positive. If not, then it is counted as a false negative. A false positive is defined as any detected onset outside all the tolerance range. 

The overall results are summarized in Table \ref{table:result}. Using the same dataset, we compared the performance of the proposed algorithm with others, including a recent algorithm proposed by Heo \emph{et al.}\cite{Heo} as well as more conventional ones [14, 1, 3, 13]. It can be seen from Figure \ref{fig:resultbar} that the proposed algorithm achieves the results significantly better than all the other algorithms in all metrics. Although not directly comparable, it is also remarkable that the performance is over 30\% higher than the best-performing algorithm for singing voice class from MIREX 2012\cite{mirex2012perclass}.

The proposed algorithm is able to detect offsets as well. However, we could not find any reliable dataset with offset annotations. Here we briefly report an extra-experiment result. An extra annotation for offsets was prepared for each one minute-long excerpt of the previous singing dataset, and for another clarinet dataset. The tolerance value for onset was set to 50 ms, as with the main experiment. On the other hand, offset tolerance was set more generously to 100 ms. In practice, human cannot find such precise defusing locations by ears. The result was $F =$ 83.5\% for singing onset and 95.0\% for clarinet onset. For offset, we obtained $F =$ 67.5\% and $F =$ 95.0\%, respectively. 
\section{Conclusion}
We proposed a pairwise approach to onset/offset detection for singing voice recordings. The proposed method differs from previous approaches in two main aspects. First, we employed higher-order statistics to capture onset/offset events in a time-domain signal. We also demonstrated that a compact adaptive kernel method improved the results. Secondly, a new peak picking algorithm was derived for this detection function. By searching a precise location where the fitness between a pre-defined kernel and the detection function is maximized, a set of onset and its corresponding offset was simultaneously found. We evaluated the proposed method with a recognized dataset. The average F-measure for onset detection by the proposed algorithm was 80.6\%,  which is highest among all methods in comparison.

\section{RELATION TO PRIOR WORK}
\label{sec:prior}
So far, a solid idea in this paper was that the higher-order statistics using correntropy \cite{Liu} would provide more robustness to general onset detection problem. A previous application to monophonic pitch detection exists\cite{Xu}. The use of {\it Rule of Thumb} for kernel parameter optimization was recommended in Liu \cite{Liu}. We extended these concepts to a novel feature representation for a detection function. 

For peak picking, adaptive threshold method has been widely used \cite{Bello}. Others have formulated such decision making into a {\it machine learning} problems \cite{Toh,Eyben,Bock}. The proposed method is novel in that it jointly estimates onset/offset, and it can be generalized as {\it dynamic programming}.

%

\vfill\pagebreak



\begin{thebibliography}{citations}

\bibitem {Bello}
J. P. Bello, L. Daudet, S. Abdallah, C. Duxbury, M. Davies and M.B. Sandler:
``A tutorial on onset detection in music signals,'' 
{\it IEEE Transactions on Speech and Audio Processing}, 
13(5), pp. 1035--1047, 2005.

\bibitem{Dixon}
S. Dixon:
``Onset detection revisited,''
In {\it Proc. of DAFx, Montreal, Canada},
pp. 133--137, 2006.

\bibitem{Duxbury}
C. Duxbury, M. Sandler, and M. Davies:
``A hybrid approach to musical note onset detection'',
In {\it Proc. of DAFx, Hamburg, Germany},
pp. 33--38,  2002.

\bibitem{Heo}
H. Heo, D. Sung and K. Lee:
``Note Onset Detection based on Harmonic Cepstrum Regularity'', 
{\it IEEE Int. Conf. on Multimedia and Expo, San Jose, USA},
pp. 1--6, 2013.

\bibitem{Toh}
C. C. Toh, B. Zhang, and Y. Wang:
``Multiple-feature fusion based onset detection for solo singing voice,''
In {\it Proc. of Int. Society for Music Information Retrieval},
2009.

\bibitem{Eyben}
F. Eyben, S. B\"{o}ck, B. Schuller and A. Graves:
``Universal onset detection with bidirectional long short-term memory neural networks,''
In {\it Proc. of Int. Society for Music Information Retrieval},
2010.

\bibitem{Bock}
S. B\"{o}ck, A. Arzt, F. Krebs and M. Schedl:
``Online real-time onset detection with recurrent neural networks,''
In {\it Proc. of DAFx},
2012.

\bibitem{Wang}
W. Wang, et. al.:
``Non-negative matrix factorization for note onset detection of audio signals,''
In {\it Proc. of IEEE Signal Processing Society Workshop on Machine Learning for Signal Processing},
2006.

\bibitem{Xu}
J. Xu and J. C. Principe:
``A pitch detector based on a generalized correlation function,''
{\it  IEEE Transactions on Audio, Speech, and Language Processing}
Vol. 16.8, pp. 1420--1432, 2008.

\bibitem{Liu}
W, Liu, P. Pokharel and J. C. Principe:
``Correntropy: properties and applications in non-Gaussian signal processing,''
{\it  IEEE Transactions on Signal Processing},
Vol. 55.11, pp. 5286--5298. 2007.

\bibitem{mirex}
J. S. Downie, A.F. Ehmann, M. Bay and M.C. Jones:
 ``The Music Information Retrieval Evaluation eXchange: Some observations and insights,''
{\it Advances in Music Information Retrieval},
Springer Berlin Heidelberg, pp. 93--115. 2010.


\bibitem{mirex2012perclass}
MIREX 2012 Onset Detection F-Measure per Class. Available from
{\url{http://nema.lis.illinois.edu/nema_out/mirex2012/results/aod/resultsperclass.html}}

\bibitem{Klapuri}
A. Klapuri: 
``Sound onset detection by applying psychoacoustic knowledge,''
In {\it Proc. of IEEE Int. Conf. Acoustics, Speech and Signal Processing, Phoenix, USA},
 pp. 115--118. 1999.

\bibitem{Schloss}
A. W. Schloss:
``On the automatic transcription of percussive music from acoustic signal to high-level analysis,''
Ph.D. thesis, Dept. Hearing and Speech, Stanford Univ., Stanford, CA, 1985, Tech. Rep. STAN-M-27.

\bibitem{mtoolbox}
O. Lartillot and P. Toiviainen:
``A Matlab toolbox for musical feature extraction from audio,''
In {\it Proc. of DAFx, Bordeaux,}
2007.

\bibitem{silverman}
B. W. Silverman, ``Density estimation for statistics and data analysis,''
London: Chapman and Hall, 1986.

\bibitem{ftest}
M. H. Kutner, C. J. Nachstheim, J. Neter and W. Li, ``Diagnostics and Remedial Measures,''
In {\it Applied Linear Statistical Models}, 5th ed. New York: McGraw-Hill/Irwin, pp. 119--124. 2005.

\bibitem{moore}
Brian. C. J. Moore and Brian R. Glasberg, ``Suggested formulae for calculating auditory filter bandwidths and excitation patterns,'' 
In {\it The Journal of the Acoustical Society of America}, 74.3, pp. 750--753. 1983.

\end{thebibliography}

\bibliographystyle{IEEEbib}

\end{document}